\documentclass[prd,preprintnumbers,floatfix,aps,notitlepage,nofootinbib,twocolumn,showpacs,amssymb]{revtex4-2}
\usepackage{amsmath,amsfonts,bm}
\usepackage{graphicx}
\usepackage{amssymb}
\usepackage{cancel}
\usepackage{booktabs} 
\usepackage{array}    
\usepackage[colorlinks, linkcolor={red},citecolor={blue}]{hyperref}
\usepackage{xcolor}

\begin{document}
\title{Kerr black holes without primary hairs}
\author{J Ovalle}
\email[]{jorge.ovalle@physics.slu.cz}
\affiliation{Research Centre for Theoretical Physics and Astrophysics,
	Institute of Physics, Silesian University in Opava, CZ-746 01 Opava,
	Czech Republic}	

\affiliation{Universidad de Tarapac\'a, Avenida Luis Emilio Recabarren 2477, Iquique, Chile
}
\begin{abstract}
	\noindent We present a class of regular axisymmetric black hole geometries fully characterized by the parameters $\{{\cal M},a\}$ and possessing the Kerr event horizon. This family interpolates between regular spacetimes, configurations with integrable singularities, and the Kerr solution as a limiting case. Its main features are: (i) the existence of quasi-extremal configurations without requiring $a \approx {\cal M}$; and (ii) a possible framework toward an analytical description of Kerr black hole formation from an initially regular configuration.

		\end{abstract} 
\maketitle
%
%
%
\section{Introduction}

\noindent Contrary to the highly idealized Schwarzschild solution, the Kerr solution~\cite{Kerr:1963ud} represents the simplest, yet minimally realistic, black hole (BH). Characterized solely by its mass and spin parameter $\{{\cal M},a\}$~\cite{Boyer:1966qh,Ernst:1967wx,Carter:1968rr,Israel:1970kp,Carter:1971zc,Robinson:1975bv}, it has been extensively studied over the years (see Refs.~\cite{Herdeiro:2014goa,Herdeiro:2015gia,Cunha:2017eoe,Gibbons:2017djb,Bambi:2019tjh,Arkani-Hamed:2019ymq,Stuchlik:2020rls,Chia:2020yla,Anson:2020trg,Afrin:2021imp,Islam:2021dyk,Cieslik:2023qdc,Chung:2023wkd,Andersson:2019dwi,Chou:2025ojv,Lo:2025njp,BenAchour:2025lkx} and references therein for representative examples of recent contributions; see also Ref.~\cite{Teukolsky:2014vca} for a comprehensive review). Nevertheless, the detailed description of the gravitational collapse leading to its formation remains an open problem. To the best of our knowledge, no simple analytical model exists that describes the emergence of a Kerr BH from an initially regular astrophysical configuration. In particular, there is no analog of the Oppenheimer-Snyder (OS) model~\cite{Oppenheimer:1939ue} capable of describing the formation of the Kerr ring singularity during the collapse process.

A development closely related to this issue is the recent formulation of an exact analytical model of gravitational collapse leading to the Schwarzschild BH~\cite{Ovalle:2025pue}. This construction describes a nonsingular Schwarzschild configuration which, unlike the OS model, possesses a highly nontrivial internal structure while remaining devoid of primary hairs. Its evolution shows, without resorting to perturbative methods~\cite{Poisson:1989zz,Ori:1991zz}, that singularities inevitably arise, in agreement with the strong cosmic censorship conjecture~\cite{Penrose:1969pc,Dafermos:2003wr,Dafermos:2017dbw,Hollands:2019whz}. It further provides, under fairly general assumptions, compelling evidence that the Schwarzschild solution cannot emerge as the continuous end-state of gravitational collapse~\cite{Ovalle:2026lxb}. Moreover, before the formation of the point source at $r=0$: (i) timelike curvature singularities necessarily develop, and (ii) Minkowski spacetime cannot be defined in a neighborhood of $r=0$, preventing the existence of a smooth local Lorentz frame at the core (see Ref.~\cite{Ovalle:2026lxb} for a detailed analysis of the near-core region and the associated ``Minkowski breaking'', Ref.~\cite{Casadio:2026tmd} for a quantum treatment of this effect, and Refs.~\cite{Lobo:2026dnl,Lobo:2026iuy} for the corresponding thermodynamic analysis).

Motivated by these results, and as a first step toward an axisymmetric analog of the OS model, we ask whether it is possible to construct a regular Kerr BH whose internal structure is sufficiently rich while remaining devoid of new charges. Following the approach of Ref.~\cite{Ovalle:2025pue}, we generate the stationary solutions by imposing: (i) weak cosmic censorship, requiring the singularity to be enclosed by an event horizon; and (ii) absence of primary hair, so that the BH is completely determined by the two quantities $\{{\cal M},a\}$, namely its total mass and angular momentum, regardless of the complexity of its internal structure. This procedure yields an infinite class of nonsingular rotating BH solutions, characterized solely by $\{{\cal M},a\}$ and of differentiability class ${\cal C}^N$ with $N>1$, whose event horizon coincides with that of Kerr. A remarkable feature of this family is the existence of quasi-extremal configurations without requiring $a\approx{\cal M}$.

 \section{Revisited Kerr BH interior}
 \label{sec1}
  \noindent Following the strategy developed in Ref.~\cite{Ovalle:2025pue}, we define a nonsingular Kerr BH configuration by (i) requiring that the spacetime belongs to the Kerr-Schild subclass throughout the region $r>0$, and (ii) recovering the Kerr vacuum solution for $r\geq\,h$, where $h$ denotes the event horizon of the Kerr BH. We begin with the Kerr-Schild metric in Boyer-Lindquist coordinates, namely the G\"urses-G\"ursey metric~\cite{Gurses:1975vu}
  \begin{eqnarray}
	\label{metric}
	ds^{2}
	&=&
	-\left[1-\frac{2\,r\,{m}(r)}{{\rho}^2}\right]
	dt^{2}
	-
	\frac{4\, {a}\, r\,{m}(r)\, \sin^{2}\theta}{{\rho}^{2}}
	\,dt\,d\phi
	\nonumber
	\\
	&&
	+
	\frac{{\rho}^{2}}{{\Delta}}\,dr^{2}
	+
	{\rho}^{2}\,d\theta^{2}
	+
	\frac{{\Sigma}\, \sin^{2}\theta}{{\rho}^{2}}\,d\phi^{2}
	\ ,
\end{eqnarray}
with
\begin{eqnarray}
	{\rho}^2
	&=&
	r^2+{a}^{2}\cos^{2}\theta\ ,
	\label{f0}
	\\
	{\Delta}
	& = &
	r^2-2\,r\,{m}(r)
	+{a}^{2}\ ,
	\label{f2}
	\\
	{\Sigma}
	& = &
	\left(r^{2}+{a}^{2}\right)^{2}
	-{\Delta}\, a^2\sin^{2}\theta\ ,
	\label{f3}
\end{eqnarray}
and
\begin{equation}
	{a}\,=\,{J}/{\cal M}\ ,
\end{equation}
where ${J}$ is the angular momentum and
\begin{equation}
	\label{cond1}
	{\cal M}\equiv m(h)
\end{equation}
is the Arnowitt-Deser-Misner (ADM) mass, where $r=h$ denotes the event-horizon radius, given by
\begin{equation}
	\label{kerrh}
	h
	=
	{\cal M}+\sqrt{{\cal M}^2-a^2}
\end{equation}
 with
\begin{equation}
	\label{mtransform}
	m(r)=\left\{
	\begin{array}{l}
		m(r)
		\ ,
		\quad
		{\rm for}\
		0< r \leq h
		\\
		\\
	{\cal M}
		\ ,
		\quad
		{\rm for}\
		r>h
		\ .
	\end{array}
	\right.
\end{equation}
 Here $m(r)$ is the Misner-Sharp mass of the spherically symmetric profile $a=0$. The classical Kerr solution is recovered from~\eqref{mtransform} by taking $m(r)={\cal M}$ for all $0<r<\infty$.
 
 To ensure smooth continuity of the metric~\eqref{metric} across $r=h$, the mass function must satisfy
 \begin{equation}
 	\label{cond2}
 	m(h)={\cal M}\ ; \qquad m'(h)=0\ ,
 \end{equation}
 where $F(h)\equiv\,F(r)\big\rvert_{r=h}$ for any $F(r)$. Condition~\eqref{cond2} guarantees that the metric is ${\cal C}^{1}$ across the horizon. 
 
 To construct an interior solution beyond the trivial case $m(r)={\cal M}$, one must adopt a generic form for $m(r)$ subject to the continuity conditions in~\eqref{cond2}. If, in addition, we demand that the interior reduces to Minkowski spacetime ($m\rightarrow\,0\,\text{as}\,\,r\rightarrow\,0$), and exhibits some form of regularity at the origin $r=0$ as measured by curvature scalars, then it becomes difficult to avoid the trivial solution $m(r)={\cal M}$, unless (i) the interior region is extended beyond the Kerr-Schild class, or (ii) additional parameters beyond $\{{\cal M},a\}$ are introduced. 
 
 Nevertheless, as we show below, it is possible to construct nontrivial solutions while remaining within the Kerr-Schild class and without introducing additional charges in the interior configuration. Although this may appear counterintuitive, a more general interior can indeed be obtained by imposing higher differentiability and requiring the mass function to be of class ${\cal C}^{N}$, as in the spherically symmetric case~\cite{Ovalle:2024wtv,Casadio:2024fol,Casadio:2025pun}. A convenient starting point is a generic solution expressed as a superposition of $N$ different contributions embedded in a de Sitter background with cosmological constant proportional to $C_3$
 \begin{equation}
	\label{mpoly}
	m(r)
	=
	C_3\,r^3
	+
	\underbrace{C_l\,r^l
		+C_n\,r^n
		+C_p\,r^p+...}_{\text{$N$ terms}}
	\ ,
\end{equation}
 where the coefficients
 \begin{equation}
 C_{s}=C_{s}({\cal M},a)\ ,
 \end{equation}
 in accordance with the premise (ii) stated in the Introduction. They are determined by imposing condition~\eqref{cond1} together with
 \begin{equation}
 	\label{cond-n}
 	\frac{d^{n} m}{dr^{n}}(h)=0\ ,
 \end{equation}
 for every $1\le n\le N$. Under these conditions, the interior mass function in~\eqref{mtransform} becomes
  \begin{eqnarray}
 	\label{minfi}
 	&&	m(r)=\left[\left(\frac{r}{h}\right)^3\,\prod_{i=1}^{N}\frac{n_i+1}{n_i-2}\right.\nonumber\\
 	&&\left.+3(-1)^N\,\sum_{k=1}^{N}\frac{1}{n_k-2}\left(\frac{r}{h}\right)^{n_k+1}\prod_{\substack{i=1\\i\neq k}}^{N}\frac{n_i+1}{n_k-n_i}\right]{\cal M}\ ,
 \end{eqnarray} 
 where $2<n_i\in\mathbb{N}$. For each fixed $N$, the solution is characterized by the set $n_i=\{n_1, n_2,\ .\ .\ .n_N\}$, which parametrizes an infinite family of regular Kerr BHs. This family depends solely on the pair $\{{\cal M},\,a\}$ of the configuration and therefore carries no primary hairs.  
 
 Regarding the roots of $\Delta(r)=0$ in~\eqref{f2}, the total number of roots is determined by the largest exponent in the set $n_i=\{n_1, n_2,\ .\ .\ .n_N\}$. Nevertheless, regardless of the value of $N$, there are always only two real roots within the region $0<r\leq\,h$. One corresponds to the event horizon $r=h$, while the other defines the inner, or Cauchy, horizon $h_c<h$. Therefore, the event horizon at $r=h$ is simple (nondegenerate). The only exception arises when the degenerate case $h_c=h$ is imposed, either through (i) the extremal condition $a={\cal M}$ or (ii) the mimicker condition (see Eq.~\eqref{mimiker} below).
 
\begin{figure*}
	\centering
	\includegraphics[width=0.325\textwidth]{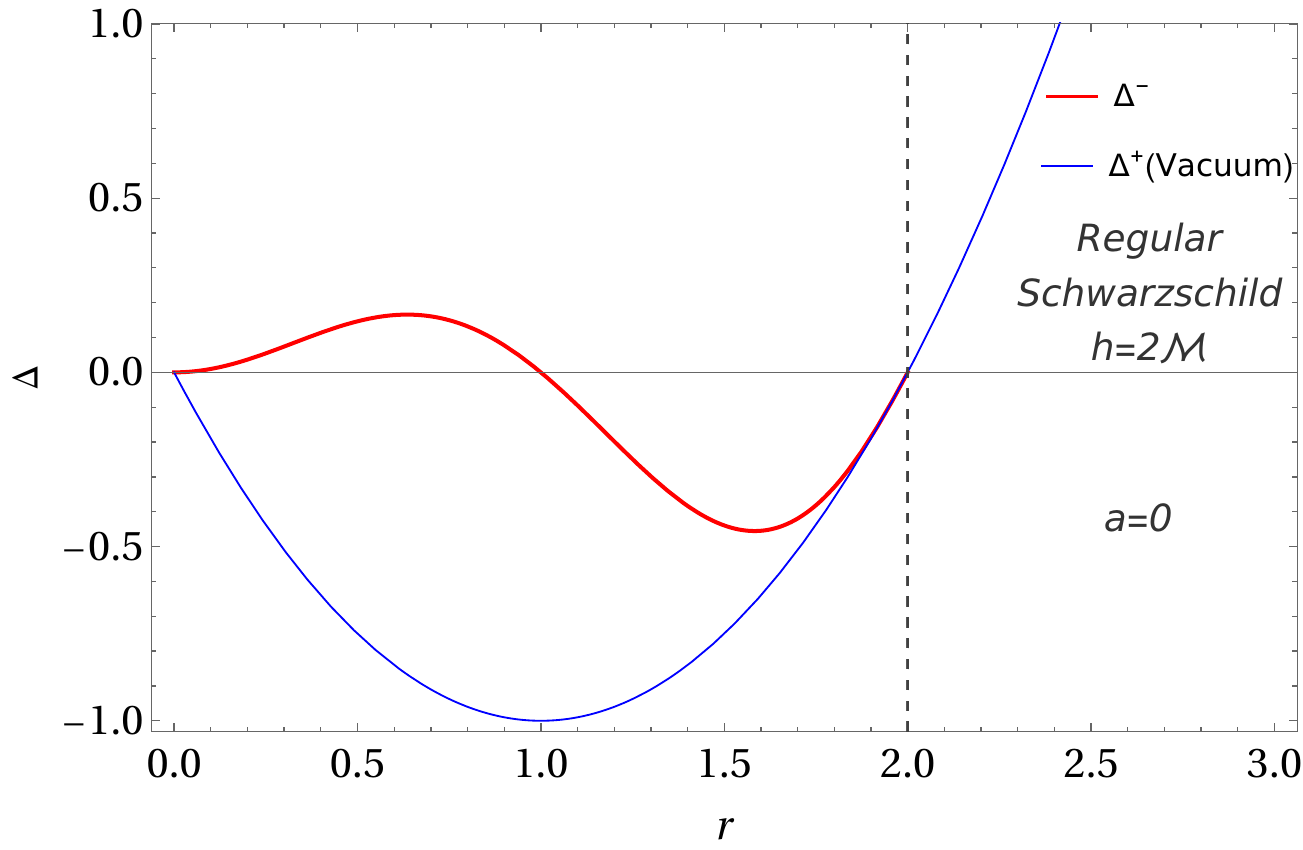} \
	\includegraphics[width=0.325\textwidth]{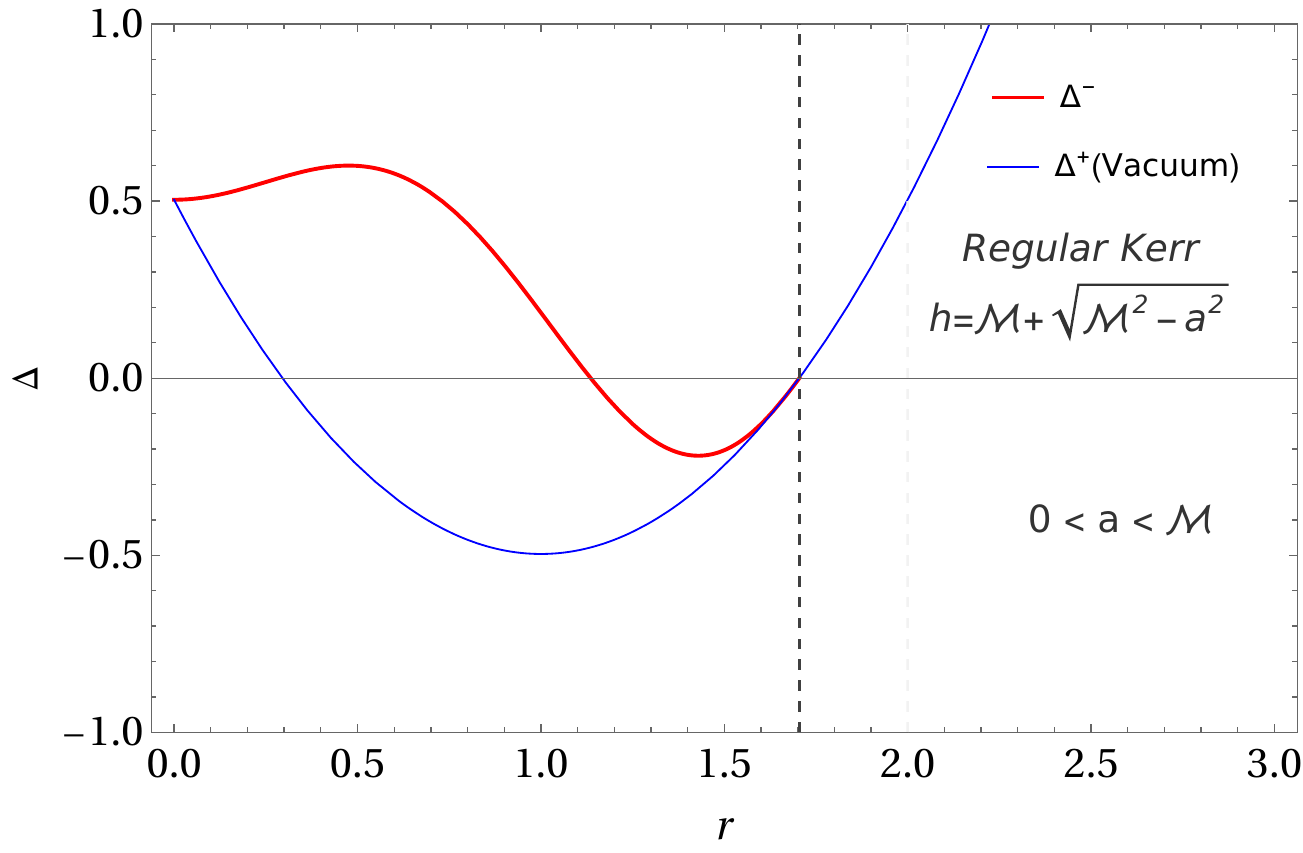} \
	\includegraphics[width=0.325\textwidth]{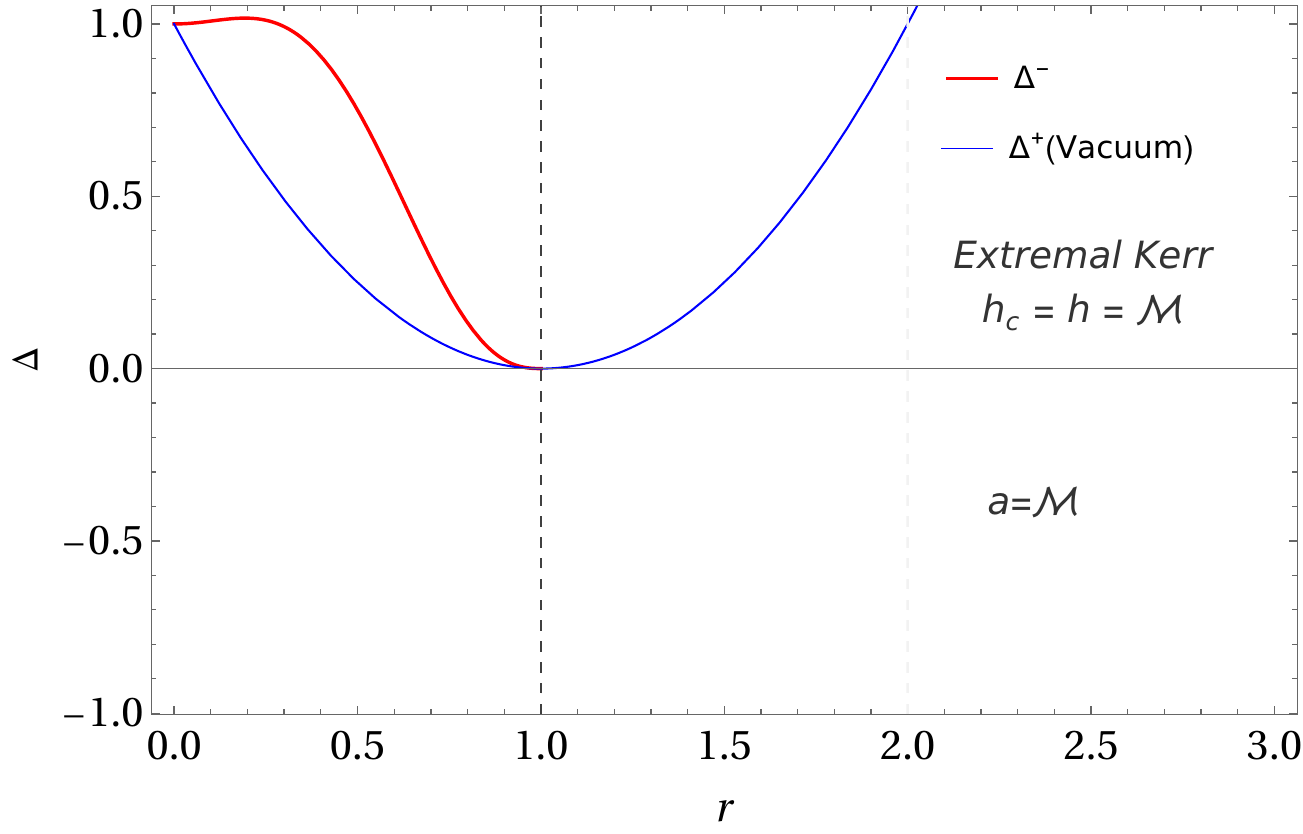} \
	\caption{\footnotesize Metric function $\Delta(r)$ in Eq.~\eqref{f2} for $N=2$ in Eq.~\eqref{minfi} [see Eq.~\eqref{m2}] with $n=3$ and $l=4$, shown for regular BH configurations with $a=0$ (Schwarzschild, left), $a\approx\,0.7{\cal M}$ (regular Kerr, center), and $a={\cal M}$ (extremal Kerr, right). For $n\ll l$, the Cauchy horizon $h_c$ lies closer to $h$. The horizon is $h={\cal M}+\sqrt{{\cal M}^2-a^2}$, and $r$ is given in units of ${\cal M}$.}
	\label{fig1}
\end{figure*}

As a concrete illustration, let us consider the simplest case with $N=1$,\footnote{In this case the term $i\neq\,k$ produces an empty product and therefore evaluates to $1$.} 
which takes the explicit form
\begin{equation}
	\label{m1}
	m(r)=\frac{{\cal M}}{(n-2)}\left[\frac{r^3}{h^3}\left(n+1\right)-3\left(\frac{r}{h}\right)^{n+1}\right]\ ;\,\,\,n>2\ ,
\end{equation}
and the case $N=2$
\begin{eqnarray}
	\label{m2}
	m(r)=&&\frac{r}{h}\left[\frac{(n+1)(l+1)}{(n-2)(l-2)}\left(\frac{r}{h}\right)^2+\frac{3\,(l+1)}{(n-2)(n-l)}\left(\frac{r}{h}\right)^n\right.
	\nonumber\\
	&&\left.+\frac{3\,(n+1)}{(l-2)(l-n)}\left(\frac{r}{h}\right)^l \right]{\cal M}\ ;\,\,\,l>n>2\,\in\mathbb{N}\ .
\end{eqnarray}
Both regular solutions~\eqref{m1} and~\eqref{m2} reduce to those originally derived in Ref.~\cite{Ovalle:2024wtv} for the spherically symmetric case ($a=0$), subsequently analyzed in Ref.~\cite{Casadio:2025pun}. See Fig~\eqref{fig1} and Table~\ref{tab} for further details.

The matter source supporting the geometry~\eqref{metric} can be written as~\cite{Burinskii:2001bq}
\begin{eqnarray}
	\label{energyax}
8\pi{\epsilon}
	&=&
	-8\pi{p}_{r}
	=
	\frac{2\,r^2}{\rho ^4}\, {m}'
	\\
	\label{pressuresax}
	8\pi{p}_{\theta}
	&=&
8\pi{p}_{\phi}
	=
	-\frac{r }{\rho ^2}\,{m}''
	+\frac{2\left(r^2-\rho^2\right)}{\rho ^4}{m}'
	\ ,
\end{eqnarray}
where ${\epsilon}$ and ${p}_i$ denote the energy density and principal pressures, respectively. Regarding the scalar curvature, it takes the form
\begin{eqnarray}
	\label{Rinfi}
	\hspace*{-6mm}	&&	R(r,\theta)=\frac{6\cal M}{\rho^2\,h}\left[4\,\prod_{i=1}^{N}\frac{n_i+1}{n_i-2}\left(\frac{r}{h}\right)^2\right.\nonumber\\
	\hspace*{-6mm}	&&\left.+(-1)^N\,\sum_{k=1}^{N}\frac{(n_k+1)(n_k+2)}{(n_k-2)}\left(\frac{r}{h}\right)^{n_k}\prod_{\substack{i=1\\i\neq k}}^{N}\frac{n_i+1}{n_k-n_i}\right]
\end{eqnarray} 
with the property that $R(h,\theta)=0$ for $N>1$. The remaining curvature invariants, such as the Kretschmann scalar, have more involved expressions, but they allow us to conclude that the ring singularity $\rho=0$ never develops as long as $n_i>2$ for all $i$. However, extending the domain to $n_i \in [{-1},2]$ yields BHs with integrable singularities~\cite{Lukash:2013ts,Ovalle:2023vvu,Arrechea:2025fkk}. In particular, if there exists an index $i$ such that $n_i=-1$, one recovers the standard Kerr solution. 

\begin{table*}
	\caption{Case $N=2$ in Eq.~\eqref{minfi} [see Eq.~\eqref{m2}]. Different choices of $\{n,l\}$ yield regular Kerr (RK) or regular Schwarzschild (RS); or extremal (E), quasi-extremal (QE) and mimicker (Mi) for both  configurations.
		\label{tab}}
	\begin{ruledtabular}
		\begin{tabular}{ c c c c c }
			$\{n,\,l\}$ & $a=0$&$0<a<{\cal M}$ &$a={\cal M}$ (except QEK)&Note
			\\
			\hline\hline
			$\{2<n<l\}$
			&
			RS, $h=2{\cal M}$& RK, $h
			=
			{\cal M}+\sqrt{{\cal M}^2-a^2}$ &EK: $h_c=h={\cal M}$ & Figure~\ref{fig1}
			\\
			\hline
			$\{2<n\ll l\}$
			&
			RS, $h=2{\cal M}$& RK, $	h
			=
			{\cal M}+\sqrt{{\cal M}^2-a^2}$  &EK: $h_c=h={\cal M}$  &Figure~\ref{fig1}
			\\
			\hline
			$\{2\ll n\ll l\}$
			&
			QES, $h_c\sim\,h=2{\cal M}$& QEK,	$h_c\sim{h}
			=
			{\cal M}+\sqrt{{\cal M}^2-a^2}$&EK: $h_c=h={\cal M}$   & ---
			\\ 
			\hline
			$\{2\ll n\ll l\}$ & 
			QES, $h_c\sim\,h=2{\cal M}$&QEK, $h_c\sim{h}
			=
			{\cal M}+\sqrt{{\cal M}^2-a^2}$ & QEK: $h_c\sim\,h\not\approx{\cal M}$  & Figure~\ref{fig3}
			\\  
			\hline
			$\{n\to\infty,l\to\infty\}$ & 
			S-Mi, $h=2{\cal M}$
			& K-Mi, $h
			=
			{\cal M}+\sqrt{{\cal M}^2-a^2}$&EK-Mi: $h={\cal M}$   & ---
			\\ 
		\end{tabular}
	\end{ruledtabular}
\end{table*}

Finally, notice that a direct inspection of the scalar curvature in Eq.~\eqref{Rinfi}, yields
the explicit expression for the effective cosmological constant
\begin{equation}
	\label{Lambda-effec}
	\Lambda_{eff}=\frac{3}{h^2}\left[\frac{2\cal M}{h}\left(\frac{r}{\rho}\right)^2\right]\prod_{i=1}^{N}\frac{n_i+1}{n_i-2}\ .
\end{equation}
We conclude by highlighting three important aspects of the metric~\eqref{metric} with the mass function~\eqref{minfi}, namely, (i) it depends only on the charges $\{{\cal M},\,a\}$ of the configuration, thus exhibiting no primary hair; (ii) it possesses a single inner (Cauchy) horizon $h_c$; and (iii) it includes the Kerr solution as a special case.
\subsection{Quasi extremal configuration }
\begin{figure}
	\includegraphics[width=0.45\textwidth]{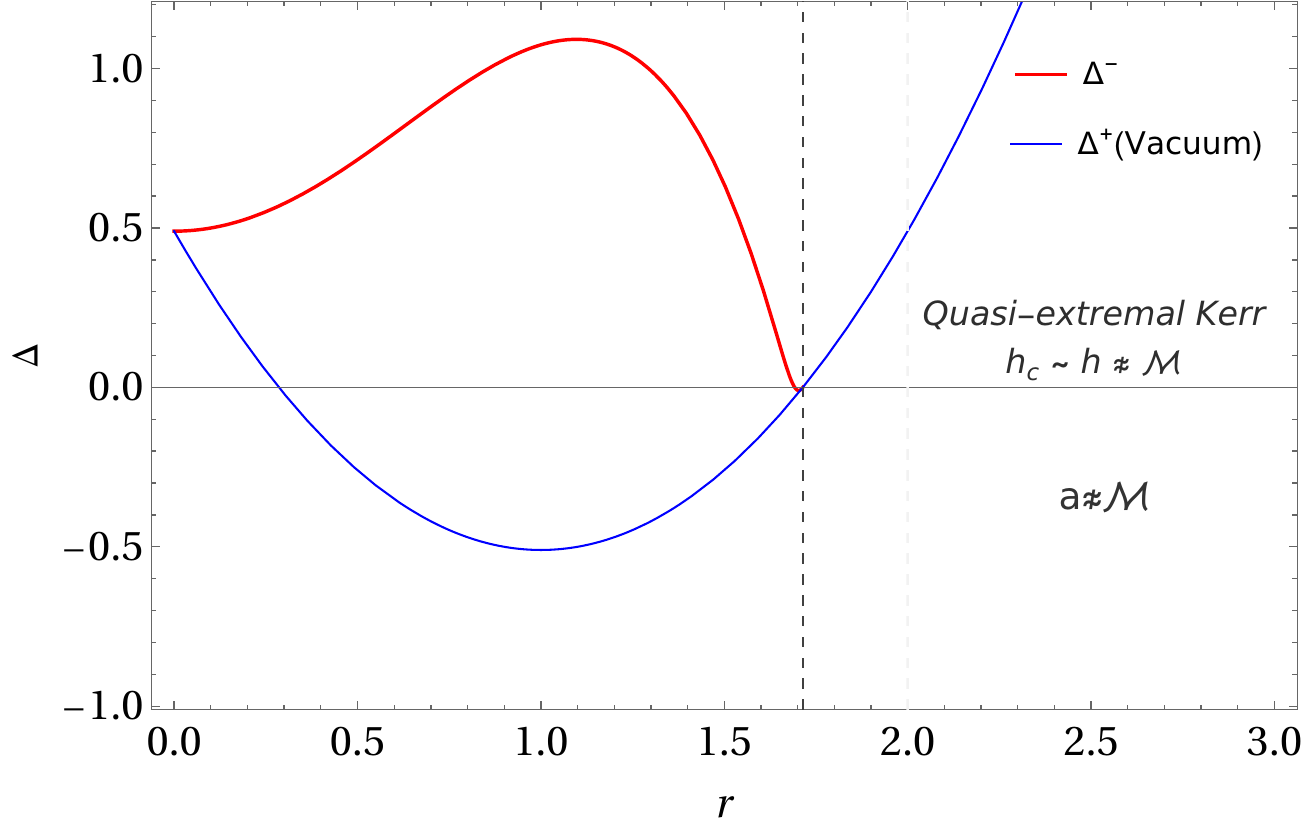} \
	\caption{\footnotesize  Metric function $\Delta(r)$ in Eq.~\eqref{f2} for $N=2$ in Eq.~\eqref{minfi} [see Eq.~\eqref{extremal}] with $n=100$, $l=200$ and $a=0.7{\cal M}$. This configuration represents a quasi extremal Kerr BH with $a\not\approx{\cal M}$ and therefore with horizon $h\not\approx{\cal M}$. The two parameters $\{{\cal M},\,a\}$ are not specially tuned. The radial coordinate $r$ is measured in units of ${\cal M}$. 
	}
	\label{fig3}
\end{figure}

\noindent We now describe a particularly interesting feature of the solutions presented above. Notice that the surface gravity $\kappa$ associated with the metric~\eqref{metric},
\begin{equation}
	\label{kappa}
	\kappa = \frac{1}{2} \frac{\Delta'(h)}{h^2+a^2} = \frac{h-{\cal M}}{h^2+a^2}\ ,
\end{equation}
remains continuous across the surface $r=h$, with the interior value $\kappa^{-}(h)$ being completely independent of the parameters $n_i$. This result shows that, regardless of the complexity of the interior structure, the horizon thermodynamics remains purely Kerr-like and is entirely determined by $h$. In contrast, the inner horizon $h_{\rm c}$, defined as a root of $\Delta(r)=0$ in~\eqref{f2}, does depend on $n_i$. Indeed, in the large-$n_i$ limit one finds $h_{\rm c}\sim h$, yielding a quasi-extremal Kerr BH with finite surface gravity, namely,
\begin{equation}
	\label{extremal}
	m(r) \to {\cal M}(r/h)^3, \quad
	h_{\rm c} \to h \quad \text{as} \quad n_i \to \infty\ \forall\,i\ .
\end{equation}
This quasi-extremal condition does not require $a \approx {\cal M}$, as illustrated in Fig.~\ref{fig3}. This result is particularly striking given that the configuration possesses no primary hair. However, in the strict limiting case
\begin{equation}
	\label{mimiker}
	m(r)={\cal M}(r/h)^3
\end{equation}
the spacetime represents a Kerr BH mimicker: it possesses a Kerr exterior, the surface $r=h$ is a null infinite-redshift surface but it is not a regular Killing horizon, since the function $\Delta(r)$ develops a cusp at $r=h$. 

\begin{table*}
	\caption{{Behavior of the Kretschmann scalar $\mathcal{K}$ for different values of the parameter $n$ associated with singular configurations, for the case $N=1$ in Eq.~\eqref{m1}. Any continuous interpolation from a regular configuration with $n>2$ to the Kerr solution at $n=-1$ necessarily crosses the Minkowski breaking (MB) at $r=0$. For $n=-2$, the Cauchy horizon $h_c$ disappears, giving rise to a spacelike strong (non-integrable) singularity.}
		\label{tab2}}
	\begin{ruledtabular}
		\begin{tabular}{ c c c c c}
			$n$ & Scaling &Notes& $\Lambda_{eff}$ &Type of Singularity
			\\
			\hline\hline
			$2$
			&
			$\mathcal{K} \sim {\cal M}^2[\log(r/h)]^2/h^6$  & $0<h_{\mathrm{c}}<h$& $\text{dS}$ &Timelike
			\\
			\hline
			$1$
			&
			$\mathcal{K} \sim {\cal M}^2(h^2r)^{-2}$   &$0<h_{\mathrm{c}}<h$ &$\text{AdS}$ &Timelike
			\\
			\hline
			$0$
			&
			$\mathcal{K} \sim {\cal M}^2(hr^2)^{-2}$    &MB&$\text{AdS}$  &Timelike
			\\ 
			\hline
			$-1$ & 
			$\mathcal{K} \sim {\cal M}^2 / r^6$
			&Kerr & $0$ &Timelike
			\\ 
			\hline
			$-2$ & 
			$\mathcal{K} \sim {\cal M}^2 h^2 / r^8$ & Cauchy horizon $h_c$ disappears& $\text{dS}$ &Spacelike
			\\  
		\end{tabular}
	\end{ruledtabular}
\end{table*}
\begin{figure}
	\centering
	\includegraphics[width=0.46\textwidth]{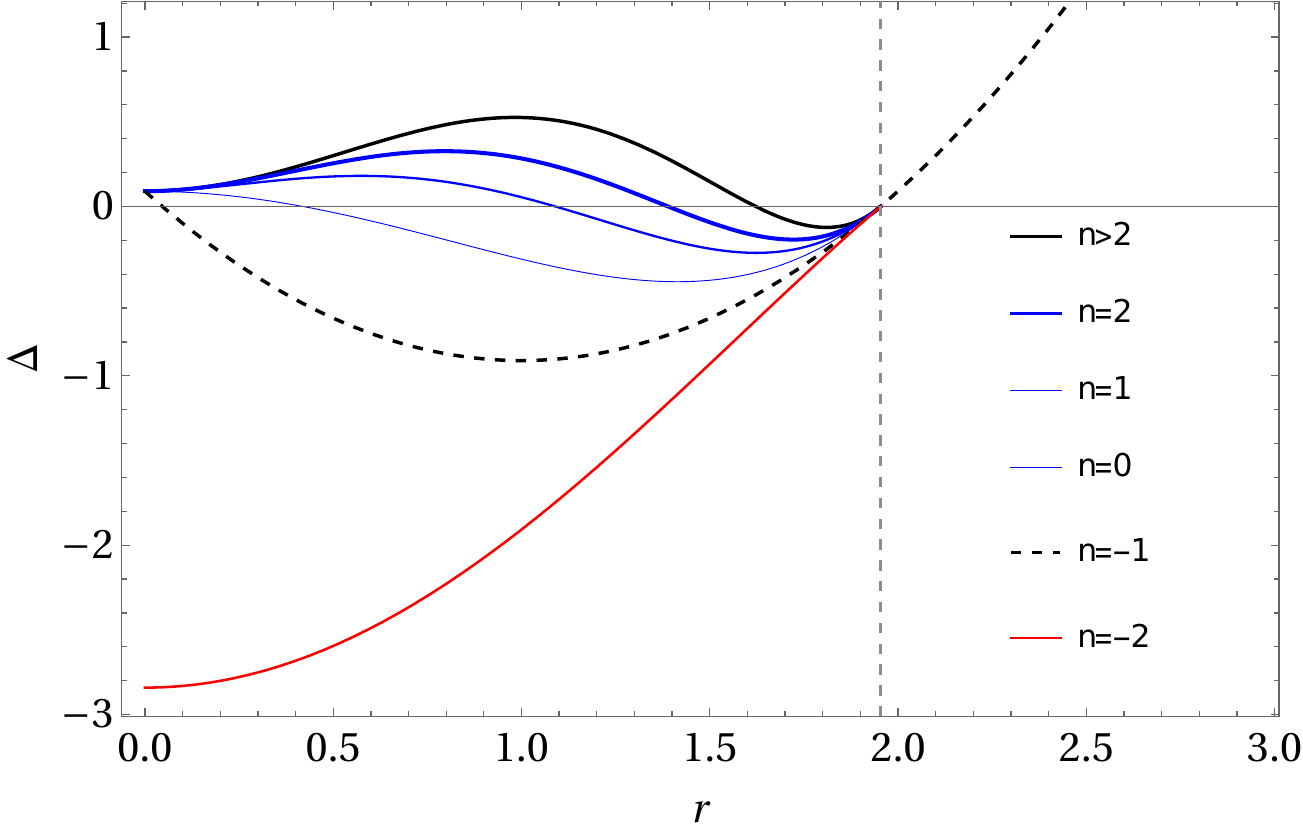} \

	\caption{\footnotesize Metric function $\Delta(r)$ in Eq.~\eqref{Delta1} for a sequence of values of $n$, interpolating between a regular configuration with $n>2$ and singular configurations as $n$ decreases (see Table~\ref{tab2}). For $n\in[-1,2]$, the configurations exhibit integrable singularities, with Minkowski breaking (MB) occurring at the core along the equatorial plane at $n=0$. The Kerr solution is recovered at $n=-1$. {We also show the case $n=-2$, where the Cauchy horizon disappears and a spacelike strong (non-integrable) singularity develops}. The radial coordinate $r$ is measured in units of ${\cal M}$.}
	\label{fig4}
\end{figure}
\section{Integrable singularities} 
\label{sec3}
\noindent As already mentioned, the geometry describes a regular black hole (RBH) whenever
\begin{equation}
	\label{RBH}
	\forall\quad i\quad n_i>2 \quad \iff\quad \text{RBH}\ .
\end{equation}
If we want to investigate the emergence of the Kerr BH, the parameter space must be extended to include values with $n_i\leq 2$. In particular, we focus on the interval $n_i \in [{-1},2]$. Spacetimes within this range correspond to BH configurations exhibiting integrable singularities (IS)~\cite{Lukash:2013ts,Ovalle:2023vvu,Arrechea:2025fkk}, that is, geometries for which the curvature scalar $R$ diverges at most as $\sim\,r^{-2}$. Hence,
\begin{equation}
	\label{IS}
	\exists\quad i\text{ such that }n_i\in [{-1},2]\iff\text{IS}\ .
\end{equation}
Within this domain, if there exists an index $i$ such that $n_i=-1$, the standard Kerr solution is recovered,
\begin{equation}
	\label{SCH}
	\exists\quad i\text{ such that }n_i=-1 \iff\text{Kerr}\ ,
\end{equation}
for which the effective cosmological constant in~\eqref{Lambda-effec} vanishes. If instead some $n_i={0}$, the Minkowski metric is not recovered as $r\rightarrow\,0$, that is, the metric is not locally $\eta_{ab}=\text{diag}(-1,1,1,1)$. This ``Minkowski breaking'' (MB) can be expressed as 
\begin{equation}
	\label{MB}
	\exists\quad i\text{ such that }n_i={0} \iff\text{MB}\ .
\end{equation}
Thus, within the integrable singularity domain, there exist a subrange $n_i \in ({0},2]$ where a locally Minkowskian frame can still be defined. {The onset of MB can be explicitly seen from the behavior of the metric component $g_{tt}$ at the core. For a given $n_i$,
\begin{equation}
	\label{MBx}
	\lim_{r \to 0} g_{tt}(r,\pi/2)=\left\{
	\begin{array}{l}
		-1
		\ ,
		\quad
		{\rm for}\
		n_i>0
		\\
		\\
		-1+3{\cal M}/h
		\ ,
		\quad
		{\rm for}\
	n_i=0
		\ .
	\end{array}
	\right.
\end{equation}
}

If instead some $n_i<{-1}$, the geometry develops strong (non-integrable) singularities (SS),
\begin{equation}
	\label{SS}
	\exists\quad i\text{ such that }n_i<{-1} \iff\text{SS}\ .
\end{equation}

Singular configurations can therefore be systematically explored by considering the stationary solutions~\eqref{minfi} arising for specific choices described by~\eqref{RBH}-\eqref{SS}. See Table~\ref{tab2} for further details.

Having identified the parameter domains associated with integrable and strong singularities, we now examine the set $n_i=\{n_1, n_2,\ .\ .\ .n_N\}$ in order to understand how such singular behavior would emerge during the collapse process. To this end, and given that $2<n_1<n_2...<n_N$, it is convenient to focus on the leading element of the ordering, namely $n_1\equiv n$. Since this parameter is the smallest, it controls the approach to the critical threshold $n=2$, where the singular behavior is triggered. Accordingly, and without loss of generality, we focus on the simplest case $N=1$ in~\eqref{m1}, which yields
\begin{equation}
	\label{Delta1}
	{\Delta}(r)=r^2+a^2-\frac{2{\cal M}r}{(n-2)}\left[\frac{r^3}{h^3}(n+1)-3\left(\frac{r}{h}\right)^{n+1}\right],
\end{equation}
while the scalar curvature~\eqref{Rinfi} takes the form
\begin{equation}
	\label{Rinfi-1}
		R(r,\theta)=\frac{6\cal M}{\rho^2\,h}\frac{(n+1)}{(n-2)}\left[4\left(\frac{r}{h}\right)^2-(n+2)\left(\frac{r}{h}\right)^{n}\right]\ .
\end{equation} 
We observe that Eqs.~\eqref{Delta1} and~\eqref{Rinfi-1} recover the Kerr solution for $n=-1$. The different configurations described by~\eqref{RBH}-\eqref{SS} are illustrated in Fig.~\ref{fig4} and summarized in Table~\ref{tab2} through the corresponding behavior of the Kretschmann scalar $\mathcal{K}$.

At this stage, one might be tempted to construct a kinematic model of gravitational collapse by promoting $n$ to a time-dependent function, $n\Rightarrow{n(t)}$, with $\dot{n}<0$, as suggested by Fig.~\ref{fig4}. In this way, one could continuously interpolate between any regular configuration with $n>2$ and the Kerr solution at $n=-1$. However, extending the stationary framework to the time-dependent case requires the construction of an exterior geometry that cannot coincide with the Kerr solution. Finding such a spacetime constitutes the main challenge in the search for an analog of the OS model for the Kerr BH, ultimately enabling a fully analytical description of its formation.

\section{Final remarks.} 

\noindent Following Penrose's singularity theorem~\cite{Penrose:1964wq,Hawking:1970zqf,Hawking:1973uf}, the infinite family of Kerr BH configurations presented in this work, described by the line element~\eqref{metric} with mass function~\eqref{minfi}, provides a natural setting in which to investigate the formation of the Kerr BH from initially regular geometries. An exact analytical description of such a process remains an outstanding challenge, in particular because there is no rotational analogue of Birkhoff's theorem.

An interior spacetime containing collapsing matter cannot, in general, be matched to an exterior described by the stationary Kerr solution. This constitutes one of the main obstacles to constructing a rotational analogue of the Oppenheimer-Snyder model, and reflects a difficulty that extends beyond those intrinsic to axisymmetry itself. The stationary configurations constructed here provide a natural geometrical framework for addressing this problem, identifying the regular configurations, the intermediate singular regimes, and the Kerr endpoint that must be connected by a time-dependent spacetime. The corresponding time-dependent construction, in which the exterior cannot remain Kerr during the evolution, is developed in the companion paper~\cite{Ovalle:2026wqw}.

\subsection*{Acknowledgments}
\vspace*{1mm}
JO thanks the Centro de Estudios Cient\'ificos (CECs) for its support during the development of this work, and in particular Ricardo Troncoso for his kind hospitality. This work was partially supported by ANID-FONDECYT Grant No. 1250227.
%

%
%
%
\bibliography{references}
\bibliographystyle{apsrev4-1}
%
%
\end{document}